\documentstyle[12pt]{article}
\begin{document}
\thispagestyle{empty}
\renewcommand{\thefootnote}{\dagger}

HURD - 9405

February, 1994.
\vskip 2truecm

\begin{center}

{\bf ASYMPTOTICALLY FINITE THEORIES AND THE SCREENING OF \\}

{\bf COSMOLOGICAL CONSTANT BY QUANTUM EFFECTS \\}

\vskip 2truecm

\bigskip
\renewcommand{\thefootnote}{\ddagger}

 {\bf I. L. Shapiro\footnote { e-mail address:
shapiro@fusion.sci.hiroshima-u.ac.jp}}

\bigskip

{\sl Tomsk State Pedagogical Institute, Tomsk, 634041, Russia \\}

{\sl and \\} {\sl Department of Physics, Hiroshima University \\}
{\sl Higashi - Hiroshima, Hiroshima, 724, Japan \\}

\end{center}
\vskip 2.5truecm

\noindent
{\bf Abstract.} In the framework of the recently proposed asymptotically
finite
gauge models the cosmological constant is essentially weakened by quantum
effects.  The next (and more general) claim is that the coupling between
quantum fields may suppress their contributions to the
induced cosmological constant.

\setcounter{page}1
\renewcommand{\thefootnote}{\arabic{footnote}}
\setcounter{footnote}0
\newpage

\section{Introduction}

The cosmological constant problem is one of the more actual ones in
a modern quantum field theory (see, for example, [1 - 7, 16 -21] and references
there). Due to the observational data we know that
the value of this constant is extraordinary small. At the same time the
appearance of the cosmological constant is
related with the vacuum energy which is generated in a variety of
quantum field theory models at various energy scales (see [1] for an
excellent review). Thus the known
theories predict the nonzero
value of the cosmological constant. In particular,
the cosmological constant arise
because of the symmetry
breaking in a
models of Grand Unification Theory (GUT)
due to the nonzero vacuum
expectation value for the Higgs bosons. The main purpose of this paper
is to
point out that for some class of GUT models there exist an effect
of the renormalization group screening of the induced cosmological
constant. The observable low-energy value
of the cosmological constant is essentualy weakened.
In fact the contributions to the cosmological constant arise within all
the realistic quantum field theory models at all energy scales.
One can suppose
that the effect described below takes
place (with a necessary modifications) at other energy scales and in
the framework of another physical models.

\section {Asymptotically finite GUT's and screening of $\Lambda$ }

In the present paper we shall restrict ourselves by consideration
of the only one kind of
GUT's with the special (asymptotically finite) kind of the renormalization
group behaviour. Such models have been proposed in [8] (see also [9 - 11]).
According to [8,11] some asymptotically finite model corresponds to every
finite
one. All the beta - functions in asymptotical finite model tend to zero
in both UV and IR limits. There is a lack of the zero - charge problem
in both limits, and
this fact is important for our purposes.
The behaviour of finite and asymptotically finite models in curved
space - time is of special interest. The action of general renormalizable
gauge model in curved space - time can be written in the form (see [9] for
the extensive review of renormalization in curved space - time):
$$
S = \int {d^4}x \sqrt{- g} \{ {\cal L}_{YM}(A_{\mu}) +
{\cal L}_{sc}({\Phi}) +
{\cal L}_{sp}({\Psi}) + {\cal L}_{int}(A, \Phi, \Psi; g,h,f) \}\eqno(1)
$$
where ${\cal L}_{YM}(A_{\mu}), {\cal L}_{sc}({\Phi}),
{\cal L}_{sp}({\Psi})$ are the Lagrangians of the Yang-Mills,
scalar and spinor fields correspondingly and
${\cal L}_{int}(A, \Phi, \Psi; g,h,f)$ is the interaction Lagrangian
which depends on the set of gauge $g$, Yukawa $h$ and scalar (Higgs)
$f$ couplings.
The Lagrangian of scalar field necessary contains the nonminimal term
${1\over2}{\xi}R{\Phi}^2$. Moreover the action of
the renormalizable theory contains
also the vacuum terms [9] but we shall not deal with these terms here.
In the framework of the asymptotically finite models there is no zero-charge
problem in UV as well as in IR region. Thus one can use the perturbation theory
in a consistent way in both regions.  In external gravitational field the
UV limit $t\rightarrow \infty$ corresponds to strong gravitational
field with large values of curvature, and IR limit $t\rightarrow
-\infty$ corresponds to weak gravitational field.
The effective couplings in the matter sector have both UV and IR fixed points
[8]. In patricular, when $t\rightarrow -\infty$ the effective constant
$f(t)$ tends to some fixed value
$f_0$. One can select some kind of asymptotycally finite theories which possess
also the property of asymptotic conformal invariance [9,11,12]. The last means
that in such theories the  effective parameters of the nonminimal
interaction $\xi(t)$ tend to the values, corresponding to the conformal
theory. On the countrary, in IR limit these couplings
grow indefinitely. It is important that
in such models $\xi(t)$ grows exponentially [13,11],
that is $\xi(t)\sim exp(- Ag^2t)$ where $A > 0$ is some model - dependent
constant, and $g$ is the fixed (nonzero) value of the gauge coupling.

Let us now pass to the cosmological constant problem. We start with the
massless
model and suppose that the scalar field mass arise as a result
of dimensional transmutation within some phase transition. Then the values
of the induced Newtonian $G$ and cosmological $\Lambda$ constants are
defined by the values of
$\langle \xi \Phi_0^2\rangle^{-1}$ and $\langle f\Phi_0^4\rangle$
where $\Phi_0$ corresponds to the minimum of the
effective potential. The effective potential can be  written as
a series in loop parameter ${\hbar} $ as
$$
V_{eff} = V + \sum^{\infty}_{n = 1}{\hbar}^n V^{(n)}   \eqno(2)
$$
where
$
V = - {1\over2} \xi R {\Phi}^2 + f {\Phi}^4
$ is classical potential of the theory (1). We consider the case
of only one scalar
field for simplicity. Since we are interesting in the low - energy
consequences of the model under consideration, we have to explore the values
of $\langle \xi \Phi_0^2\rangle^{-1}$ and $\langle f\Phi_0^4\rangle$
in the  IR limit $t\rightarrow -\infty$.
For these purposes we shall use the general solution of the
renormalization group
equation for effective potential in curved space-time
(see, for example, [9]).
$$
V_{eff}[e^{-2t}g_{\alpha\beta},{\Phi},f,\xi ,\mu ] =
e^{-4t}V[g_{\alpha\beta},{\Phi}(t),f(t),\xi(t),\mu ]
\eqno(3)
$$
Here $\mu$ is dimensional parameter of renormalization. The effective
coulings and fields satisfy the renormalization group equations
of the form
$$
{d\Phi (t)\over{dt}} = ({\gamma}_{\Phi} - 1)\Phi
,\;\;\;\;\;\;\;\;\;
\Phi (0) = \Phi   $$$${df(t)\over{dt}} = {\beta} _f,\;\;\;\;\;\;\;\;\;\;\;
f(0) = f  $$$${d\xi (t)\over{dt}} = {\beta}_{\xi}
,\;\;\;\;\;\;\;\;\;\;\;\;
\xi (0) = \xi
\eqno(4)
$$

Generally speaking the exploration of the IR asymptotics of $G$ and $\Lambda$
needs the research of the corresponding limits of $V_{eff}$.
At the same time one can obtain the general answer in a more simple way.
Let us consider the zero order approximation for $V_{eff}$ taking the
renormalization group improved classical potential $V$. Then we have to
substitute the fields and couplings in the classical potential by the
effective ones (4) according to (3).
So we find
$$
G^{-1} \sim  \langle \xi (t) \Phi_0^2 (t) \rangle
\eqno(5a)
$$
$$
{\Lambda \over {G}} \sim -\langle f(t)\Phi_0^4 (t) \rangle
\eqno(5b)
$$

Note that the beta - functions (4) are well - defined, unlike ${\gamma}_\Phi$
which is not.
${\gamma}_{\Phi}$ contains an essential
arbitrariness, related with the dependence of the choice of gauge in the
Yang - Mills sector and also with the arbitrariness in the parametrization
of quantum fields. Therefore the only equation (4) for $\Phi$ is
insufficient for the determination of the asymptotic of this effective
charge. At the same time we can obviously find this asymptotic from $(5a)$.
In fact, since $G(- \infty)$ have the finite classical value
and
 $\xi(t)\sim exp(- Ag^2t)$ we find that
 $\Phi(t)\sim exp(Ag^2t)$ .
Now we take into account (5b). Since in the asymptotically finite
models $\lambda (t) \rightarrow {\lambda}_0 = const$ we find that in
 the IR limit $t \rightarrow -\infty$
$$
\Lambda \sim \langle f(t) \Phi_0^4 (t) \rangle
\sim exp(2Ag^2) \rightarrow 0
\eqno(6)
$$
So we have found that in given approximation the cosmological constant
decrease exponentially simultaneously with the decrease of the energy
scale, and hence the  observable value  of $\Lambda$
will be essenially weakened.

In the above consideration we have used the renormalization group improved
classical potential.
In fact the higher loop contributions have the structure
which differs from $V$ by some terms containing $ln\Phi$. Since these terms
have the softer IR behaviour in comparison with ${\Phi}^4$ or ${\Phi}^2$
one can obviously wait for just the same asymptotic
behaviour of $V_{eff}$ when the loop corrections are taken into account.

\section {Some attempt of numerical estimate}

 Now we can try to make some numerical estimate of the effect considered
above. Suppose that
the asymptotically finite model describe the early Universe between some energy
scales ${\mu }_{IR}$ and ${\mu}_{UV}$.
Then from (4),(5b) and (6) we obtain the following relation for the ratio
$$
{{\Lambda}_{IR} \over {\Lambda}_{UV}}
= {({{\mu}_{IR} \over {{\mu}_{UV}}})}^{2Ag^2}
\eqno(7)
$$
Hence the result of our estimate essentially depends on the choice of the
model (that define $A$) as well as on the region of it's application.
One can make, for instance, the following "optimistic" estimate. Let, for
instance, the
upper bound corresponds to Planck energies ${\mu}_{UV} = M_p
\approx 10^{19}Gev$ and the lower
bound - to the energy of the relic radiation ${\mu}_{IR} = M_r \approx 10^{-12}
Gev$. The values of $A$ have been calculated in a number of papers (see, for
example, [13,12,11] and [9] for finite and asymptotically finite models). Here
we shall use the results of [11], where the bounds for $A$ are $5 < A < 50$.
Taking the optimistic value $A = 50$ we get
$$
{{\Lambda}_{IR} \over {\Lambda}_{UV}}
\approx 10^{- 3100g^2}     \eqno(8)
$$
Therefore if one suppose that $g\approx 10^{-1}$ then the value of $\Lambda$
is decreased on 31 order.

One can truly find above estimate as very naive. In fact we can not
 wait that some GUT model is valuable at more than 10 orders of the energy
scale. Then we have to substitute 3100 in (8) by 1000 or even a less number.
However there is another way to enlarge the power in
(7). Generally speaking, $A$ is constant only in finite models, but not in
asymptotically finite ones. If one consider the asymptotically
finite theory at the middle
          (not UV or IR) character energies, the values of scalar
and Yukawa couplings may differ from the fixed ones [8]. If one choose the
large
enough value for $f(0)$ then the coefficient $A$ will be larger
(see [9] for the details of the renormalization group equation for $\xi$)
and the
considered effect will give more essential quantative improvements
for the value of $\Lambda$ at low energies.

\section {Discussion}

We have considered the renormalization group mechanism of suppression of the
cosmological constant. Some notes are in order. The described scheme
doesn`t need the fine tuning of the parameters of the theory. However
if one want to deal with the asymptotically finite theories, the rigid
resrictions on the multiplet composition takes place (see [8,11] and
references there for this point). One - loop models of
this type have been proposed in [11] (see also [14] for some generalization
concerning the finiteness in a massive sector). The direct use of the
above scheme in higher loops needs the finiteness in the corresponding
order of perturbation theory
that is possible only in supersymmetric theories (see [15] and references
there).
However the renormalization groop screening of $\Lambda$ may take place
in the non - finite models as well. In fact the more essential thing
is a large enough value of the coupling $g$  and the choice of the model
with large (and positive!) $A$ in (7).
Therefore it is natural to suppose that the
effective screening of $\Lambda$ is
possible in an arbitrary kind of models with strong coupling.

The discussed mechanism is based on the assumption that the necessary
phase transition really takes place, and that the finite value of the
 gravitational constant is induced.
Since we relate the values of induced constants $G$ and $\Lambda$ with the IR
asymptotics of
$\langle \xi \Phi_0^2\rangle$ and $\langle f\Phi_0^4\rangle$
it should be very interesting to explore the IR asymptotic behaviour of
the corresponding composite operators, and so to get a more rigid prove
for the screening of $\Lambda$.

\section {Acknowledgments}

Author is grateful to I.L.Buchbinder and I.V.Tyutin for helpful discussions.

 \end{document}